\documentclass[aps,prb,a4paper,10pt,twocolumn,showpacs,floatfix,bibnotes,superscriptaddress,preprintnumbers,longbibliography]{revtex4-2}
\usepackage{float}
\usepackage{dcolumn,graphicx,color,booktabs,microtype,afterpage}
\usepackage{amssymb}
\usepackage{amsmath}
\usepackage[charter,greekuppercase=italicized]{mathdesign}
\usepackage{sidecap}
\usepackage[mathlines]{lineno}

\graphicspath{{./}{figure/}}
\renewcommand{\tablename}{Table}
\makeatletter\renewcommand{\fnum@figure}[1]{\figurename~\thefigure.~}\makeatother
\makeatletter\renewcommand{\fnum@table}[1]{\tablename~\thetable.}\makeatother

\newcount\hh \newcount\mm
\hh=\time \divide\hh by 60
\mm=\hh \multiply\mm by 60 \mm=-\mm
\advance\mm by \time
\def\now{\number\hh:\ifnum\mm<10{}0\fi\number\mm}

\usepackage[colorlinks,plainpages=false,linkcolor=blue,urlcolor=blue,citecolor=blue,pdfpagemode=UseNone,pdfstartview=FitBH]{hyperref}
\usepackage{nicefrac}

\newcommand{\tcr}[1]{\textcolor{black}{#1}}

\begin{document}

\makeatletter\renewcommand{\ps@plain}{%
\def\@evenhead{\hfill\itshape\rightmark}%
\def\@oddhead{\itshape\leftmark\hfill}%
\renewcommand{\@evenfoot}{\hfill\small{--~\thepage~--}\hfill}%
\renewcommand{\@oddfoot}{\hfill\small{--~\thepage~--}\hfill}%
}\makeatother\pagestyle{plain}

\preprint{\textit{Preprint: \today, \now}} 

\title{Fully-gapped superconductivity and topological aspects of \\
	the noncentrosymmetric TaReSi superconductor}
\author{T.\ Shang}\email[Corresponding authors:\\]{tshang@phy.ecnu.edu.cn}
\affiliation{Key Laboratory of Polar Materials and Devices (MOE), School of Physics and Electronic Science, East China Normal University, Shanghai 200241, China}
\affiliation{Chongqing Key Laboratory of Precision Optics, Chongqing Institute of East China Normal University, Chongqing 401120, China}
\author{J.\ Z.\ Zhao}\email[Corresponding authors:\\]{jzzhao@swust.edu.cn}
\affiliation{Co-Innovation Center for New Energetic Materials, Southwest University of Science and Technology, Mianyang 621010, China} 
\author{Lun-Hui\ Hu} 
\affiliation{Department of Physics and Astronomy, University of Tennessee, Knoxville, Tennessee 37996, USA}
\author{D.\ J.\ Gawryluk}
\affiliation{Laboratory for Multiscale Materials Experiments, Paul Scherrer Institut, CH-5232 Villigen PSI, Switzerland}

\author{X.\ Y.\ Zhu}
\affiliation{Key Laboratory of Polar Materials and Devices (MOE), School of Physics and Electronic Science, East China Normal University, Shanghai 200241, China}
\author{H.\ Zhang}
\affiliation{Key Laboratory of Polar Materials and Devices (MOE), School of Physics and Electronic Science, East China Normal University, Shanghai 200241, China}
\author{J.\ Meng}
\affiliation{Key Laboratory of Polar Materials and Devices (MOE), School of Physics and Electronic Science, East China Normal University, Shanghai 200241, China}

\author{Z.\ X.\ Zhen}
\affiliation{Key Laboratory of Polar Materials and Devices (MOE), School of Physics and Electronic Science, East China Normal University, Shanghai 200241, China}
\author{B.\ C.\ Yu}
\affiliation{Key Laboratory of Polar Materials and Devices (MOE), School of Physics and Electronic Science, East China Normal University, Shanghai 200241, China}

\author{Z.\ Zhou}
\affiliation{Key Laboratory of Nanophotonic Materials and Devices \&
	Key Laboratory of Nanodevices and Applications,
	Suzhou Institute of Nano-Tech and Nano-Bionics (SINANO), CAS, Suzhou 215123, China}
\author{Y.\ Xu}
\affiliation{Key Laboratory of Polar Materials and Devices (MOE), School of Physics and Electronic Science, East China Normal University, Shanghai 200241, China}
\author{Q.\ F.\ Zhan}
\affiliation{Key Laboratory of Polar Materials and Devices (MOE), School of Physics and Electronic Science, East China Normal University, Shanghai 200241, China}
\author{E.\ Pomjakushina}
\affiliation{Laboratory for Multiscale Materials Experiments, Paul Scherrer Institut, CH-5232 Villigen PSI, Switzerland}
%
%
\author{T.\ Shiroka} 
\affiliation{Laboratory for Muon-Spin Spectroscopy, Paul Scherrer Institut, CH-5232 Villigen PSI, Switzerland}
\affiliation{Laboratorium f\"ur Festk\"orperphysik, ETH Z\"urich, CH-8093 Z\"urich, Switzerland}

\begin{abstract}
We report a study of the noncentrosymmetric TaReSi superconductor by 
means of muon-spin rotation and relaxation ({\textmu}SR) technique, 
complemented by electronic band-structure calculations. 
Its superconductivity, with $T_c$ = 5.5\,K and 
upper critical field $\mu_0H_\mathrm{c2}(0)$ $\sim$ 3.4\,T, was 
characterized via electrical-resistivity- and magnetic-susceptibility measurements.
The temperature-dependent superfluid density, obtained from
transverse-field {\textmu}SR, suggests a fully-gapped superconducting state in TaReSi, with an energy gap 
$\Delta_0$ = 0.79\,meV and a magnetic penetration depth $\lambda_0$ = 562\,nm.
The absence of a spontaneous magnetization below $T_c$, as confirmed by zero-field {\textmu}SR, indicates a preserved
time-reversal symmetry in the superconducting state. 
The density of states near the Fermi level is dominated by the Ta- and 
Re-5$d$ orbitals, which account for the relatively large band 
splitting due to the antisymmetric spin-orbit coupling. 
In its normal state, TaReSi behaves as a three-dimensional
Kramers nodal-line semimetal, characterized by an hourglass-shaped
dispersion protected by glide reflection. By combining
non\-triv\-i\-al electronic bands with intrinsic superconductivity, 
TaReSi is a promising material for investigating the topological
aspects of noncentrosymmetric superconductors.
\end{abstract}

\maketitle\enlargethispage{3pt}

\vspace{-5pt}
\section{\label{sec:Introduction}Introduction}\enlargethispage{8pt}
In crystalline solids, a suitable combination of space-,
time-reversal-, and parity symmetries often gives rise to exotic 
quasiparticles, analogous to the particles predicted in high-energy 
physics, such as Dirac-, Weyl-, or Majorana fermions~\cite{Armitage2018,Lv2021,Yan2017,Wieder2022,Bradlyn2017,Vergniory2019,Vergniory2019,Tang2019,Bernevig2022}.
In particular, materials which lack an inversion center are among 
the best candidates for studying topological phenomena, since most of them also exhibit  nonsymmorphic symmetry that
can often generate unusual 
types of fermionic excitations. 
For instance, Weyl fermions were experimentally discovered as quasiparticles
in noncentrosymmetric tantalum- and niobium pnictides~\cite{Xu2015a,Xu2015b,Lv2015,Xu2016}. 
Noncentrosymmetric materials can 
also host exotic fermions with an 
hourglass-shaped dispersion protected by glide 
reflection~\cite{Wang2016,Wu2019,Wu2020}, which are 
known to exhibit interesting topological properties. In addition, 
Kramers nodal-line (KNL) fermions were recently forecasted 
to occur in noncentrosymmetric metals with a sizable
spin-orbit coupling (SOC)~\cite{Chang2018,Xie2021}.  	
To date, research on topological materials has been primarily focused on the 
case of non-interacting electronic bands. 
On the contrary, the interplay between 
topology and correlated electronic states, such as superconductivity 
or magnetism, remains largely unexplored.

Many noncentrosymmetric topological materials
also exhibit superconductivity (SC) and, in view of their structure, are known as noncentrosymmetric superconductors (NCSCs). 
In NCSCs, the antisymmetric spin-orbit coupling (ASOC) allows, in principle, the occurrence of admixtures of spin-singlet
and spin-triplet superconducting pairing, whose degree of mixing is generally believed to be determined by the strength of ASOC~\cite{Bauer2012,Smidman2017,Ghosh2020b}.
This sets the scene for a variety of exotic superconducting properties, e.g.,  
nodes in the energy gap~\cite{yuan2006,nishiyama2007,bonalde2005CePt3Si,Shang2020}, 
multigap SC~\cite{kuroiwa2008}, upper critical fields beyond the Pauli 
limit~\cite{Carnicom2018,Bauer2004,Su2021}, and breaking of time-reversal 
symmetry (TRS) in the superconducting state~\cite{Shang2020,Hillier2009,Barker2015,Shang2020b,Singh2014,Shang2018a,Shang2018b,Shang2020ReMo}.

Noncentrosymmetric superconductors also provide a fertile ground 
in the search for topological SC and Majorana zero modes, with potential applications to 
quantum computation~\cite{Sato2017,Qi2011,Kallin2016,Kim2018,Sun2015,Ali2014,Sato2009,Tanaka2010}. 
Among the many routes attempted to realize it, one approach 
consists in combining a conventional $s$-wave superconductor
with a topological insulator to form a heterostructure. 
The proximity effect between the resulting surface states can 
lead to an effective two-dimensional SC with $p$+$ip$ pairing, known to support 
Majorana bound states at the vortices~\cite{Fu2008,Xu2014,Peng2014}.
One can also consider introducing extra carriers (e.g., via chemical doping) into a topological insulator to achieve topological superconductivity~\cite{Hor2010,Sasaki2011}.
A more elegant
and clean route to attain topological SC is that of 
combining a nontrivial electronic band with intrinsic superconductivity in the same compound~\cite{Zhang2018}.
\tcr{Some of the materials with nontrivial electronic band structures display 
	topological surface states with spin-polarized textures~\cite{Armitage2018,Lv2021,Yan2017,Wieder2022,Bradlyn2017,Vergniory2019,Vergniory2019,Tang2019,Bernevig2022}.}
When the bulk of the material transitions into the 
superconducting state, the proximity effect can give rise to 
topological superconducting surface states. 
Such protected surface states have been
proposed, for instance, in noncentrosymmetric $\beta$-Bi$_2$Pd and PbTaSe$_2$ superconductors, 
both considered as suitable
platforms for investigating topological SC~\cite{Guan2016,Sakano2015}.
Clearly, to pursue the ``intrinsic'' route, 
it is of fundamental interest to identify new types 
of superconductors with a nontrivial band topology. 

\tcr{Recently, NSCSs have become one of the most investigated superconducting
classes due to their unconventional- and topological nature.}
\tcr{To this superconducting family belong also the TiFeSi-type materials,
such as $T$RuSi and $T$ReSi (with $T$ a transition metal). 
The normal states of TaRuSi and NbRuSi are three-dimensional KNL semimetals, characterized by large ASOCs and by hourglass-like dispersions~\cite{Shang2022b}. 
Both compounds spontaneously break the TRS in the superconducting state
and adopt a unitary ($s + ip$) pairing, reflecting a mixture of spin
singlets and spin triplets. 
%
%
\begin{figure}[!thp]
	\centering
	\vspace{-1ex}%
	\includegraphics[width=0.44\textwidth,angle=0]{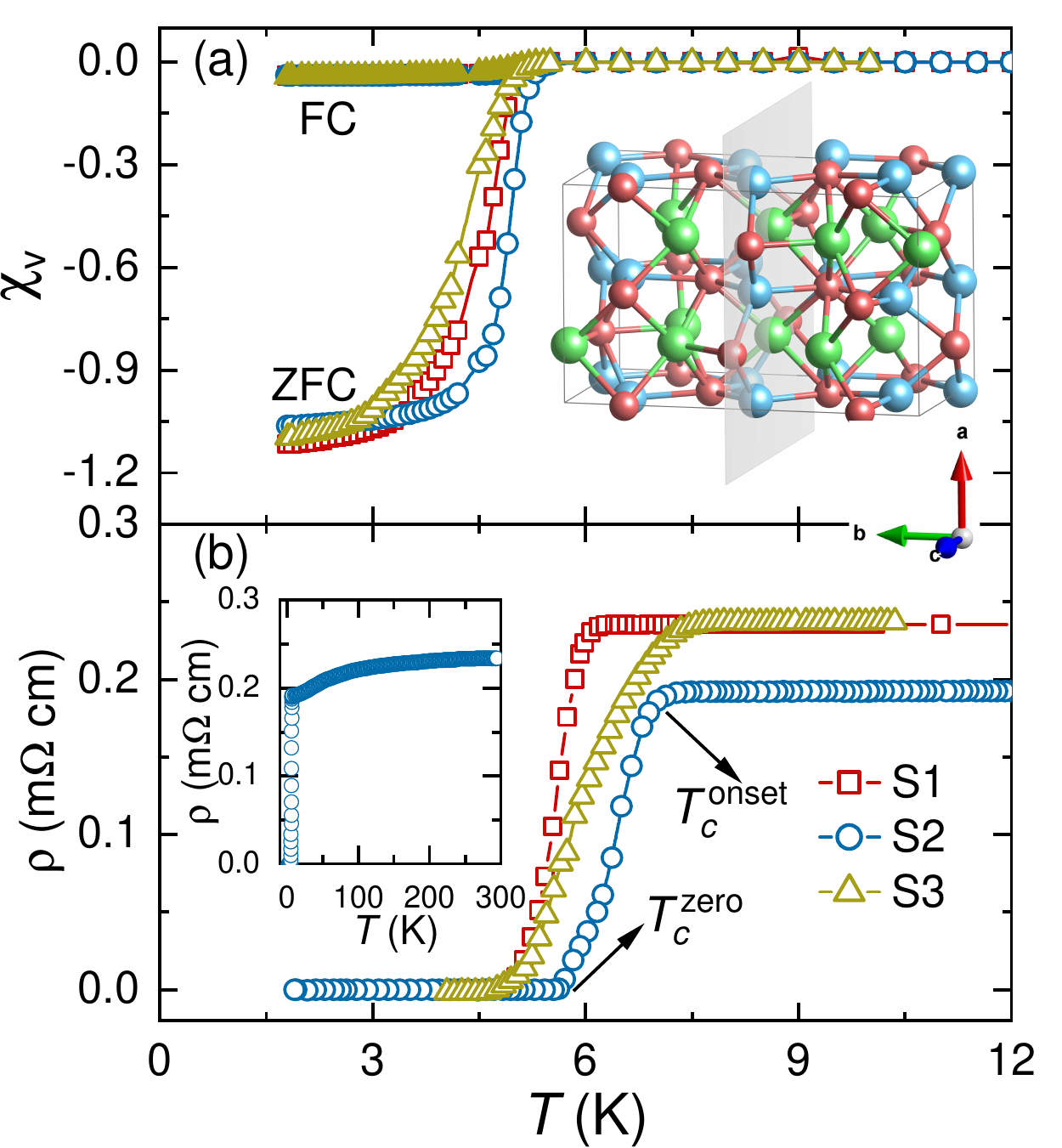}
	\caption{\label{fig:Tc}
	The temperature-dependent volume magnetic susceptibility 
	$\chi_\mathrm{V}(T)$ (a) and electrical resistivity 
	$\rho(T)$ (b) for TaReSi. 
	The results of an as-cast sample (S1) and of samples annealed 
	at 900$^\circ$C (S2) and 1100$^\circ$C (S3) are shown. 
	While $\rho(T)$ was measured in a zero-field condition,
	$\chi_\mathrm{V}(T)$ data were collected in a magnetic field of
	1\,mT. The susceptibility data were corrected to account for the
	demagnetization factor.  The inset in (a) shows the crystal structure
	of TaReSi viewed along the $c$-axis and its mirror plane, 
	while the black lines mark the unit cell. 
	Green, blue, and red spheres are Ta, Re, and Si atoms, respectively. 
    The inset in (b) shows the $\rho(T)$ of S2 up to 300\,K.}
\end{figure}
%
TaReSi also belongs to the TiFeSi family, and becomes a superconductor
below 5.5~K~\cite{Subba1985}.}
Although certain properties of TaReSi have been previously investigated~\cite{Sajilesh2021}, its superconducting properties, in particular, the superconducting order parameter, have not been explored at a microscopic level.
\tcr{In this paper, by combining muon-spin relaxation and rotation ({\textmu}SR) measurements with electronic band-structure calculations,
we show that TaReSi exhibits a fully-gapped superconducting state with a preserved TRS. 
It shares similar band topology with TaRuSi and NbRuSi, whose Kramers-
and hourglass fermions can be easily tuned towards the Fermi level by chemical substitutions.  TaReSi serves as another candidate material for 
investigating the interplay between topological states and superconductivity.}
%

\vspace{-2.5ex}
\section{Experimental details\label{sec:details}}\enlargethispage{8pt}
Polycrystalline TaReSi samples were prepared by arc melting stoichiometric 
Ta slugs (Alfa Aesar, 99.98\%), Re powders (ChemPUR, 99.99\%), and Si chunks 
(Alfa Aesar, 99.9999\%) in a high-purity argon atmosphere. To improve 
sample homogeneity, the ingots were flipped and re-melted 
more than six times. The resulting samples were then separated and 
annealed at 900$^\circ$C and 1100$^\circ$C for two weeks, respectively. 
An as-cast sample (denoted as S1) and samples annealed at 
900$^\circ$C (S2) and 1100$^\circ$C (S3) were studied. 
As shown in the inset of Fig.~\ref{fig:Tc}, TaReSi crystallizes in an 
orthorhombic structure with a space group of $Ima2$ (No.\,46)~\cite{Sajilesh2021}. 
All samples were characterized by electrical-resistivity- 
and magnetization measurements, 
performed on a Quantum Design physical property measurement 
%
%
\begin{figure}[!thp]
	\centering
	\vspace{-1ex}%
	\includegraphics[width=0.47\textwidth,angle=0]{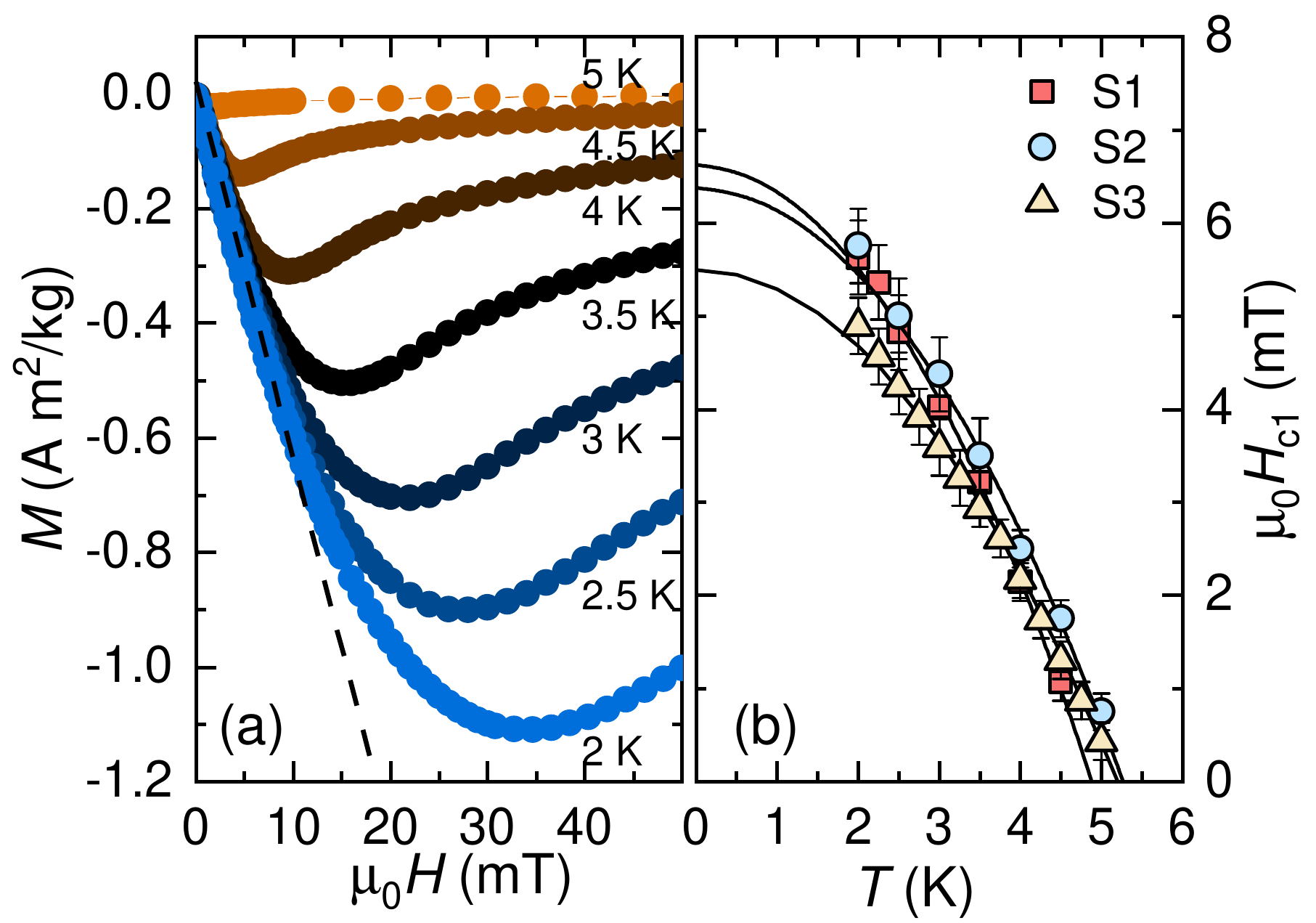}
	\caption{\label{fig:Hc1}(a) Field-dependent magnetization curves
	collected at various temperatures after cooling the S2 
	sample in zero field (the other samples behave similarly).
	(b) Lower critical fields $H_\mathrm{c1}$ vs.\ temperature. Solid lines are fits to $\mu_{0}H_\mathrm{c1}(T) =\mu_{0}H_\mathrm{c1}(0)[1-(T/T_{c})^2]$.
	For each temperature, $H_\mathrm{c1}$ was determined as the 
	value where $M(H)$ starts deviating from linearity (see dashed line).}
\end{figure}
%
%
system (PPMS) and a magnetic property measurement system (MPMS), respectively.
The bulk {\textmu}SR measurements were carried out at the multipurpose 
surface-muon spectrometer (Dolly) on the $\pi$E1 beamline of the Swiss muon source at Paul 
Scherrer Institut, Villigen, Switzerland. The samples were mounted on 
a 25-{\textmu}m thick copper foil which ensures 
thermalization at low temperatures.
The time-differential {\textmu}SR data were collected upon heating and 
then analyzed by means of the \texttt{musrfit} software package~\cite{Suter2012}.

First-principles calculations were performed based on the density functional theory (DFT), as implemented in the \texttt{Quantum ESPRESSO} package~\cite{giannozzi2009,giannozzi2017}.
The exchange-correlation function was treated with the generalized gradient approximation using the Perdew-Burke-Ernzerhof (PBE) realization~\cite{Perdew1996iq}.
The projector augmented wave pseudopotentials were adopted~\cite{Blochl1994zz}.
We considered 13 electrons for Ta (5$s^{2}$6$s^{2}$5$p^6$5$d^{3}$), 
15 electrons for Re (5$s^{2}$6$s^{2}$5$p^6$5$d^{5}$), and 
4 electrons for Si (3$s^{2}$3$p^{2}$) as valence electrons. 
The calculations use the measured lattice parameters $a = 7.002$~\AA{}, $b = 11.614$~\AA{}, and
$c = 6.605$~\AA{}, and coordinates  Ta$_1$ (0.2500, 0.2004, 0.2964), Ta$_2$ (0.2500, 0.7793, 0.2707), Ta$_3$ (0.2500, 0.9979, 0.9178),  Re$_1$ (0, 0, 0.25), Re$_2$ (0.0295, 0.3764, 0.1200), and Si$_1$ (0.25, 0.9747, 0.5055), Si$_2$ (0.0060, 0.1675, 0.9953) for TaReSi and include also the spin-orbit coupling effects~\cite{Rao1985}.
The kinetic energy cutoff for the wavefunctions was set to 60 Ry, 
while for the charge density it was fixed to 600 Ry. 
For the self-consistent calculations, the Brillouin zone integration 
was performed on a Monkhorst-Pack grid mesh of $10 \times 10 \times 10$ 
$k$-points, which ensures their unbiased sampling.
The convergence criterion was set to $10^{-7}$ Ry.
The Hf- and W doping effects were simulated by the virtual
crystal approximation (VCA)~\cite{bellaiche_virtual_2000} implemented
in the Vienna Ab initio Simulation Package (VASP)~\cite{Kresse1996kl,Kresse1996vk}.

\vspace{-2ex}
\section{Results and discussion\label{sec:results}}\enlargethispage{8pt}

The bulk superconductivity of the TaReSi samples was
first characterized by magnetic-susceptibility measurements, using both 
field-cooling (FC) and zero-field-cooling (ZFC) protocols
in an applied magnetic field of 1\,mT. As shown in Fig.~\ref{fig:Tc}(a), 
\begin{figure}[!htp]
	\centering
	\includegraphics[width=0.46\textwidth,angle= 0]{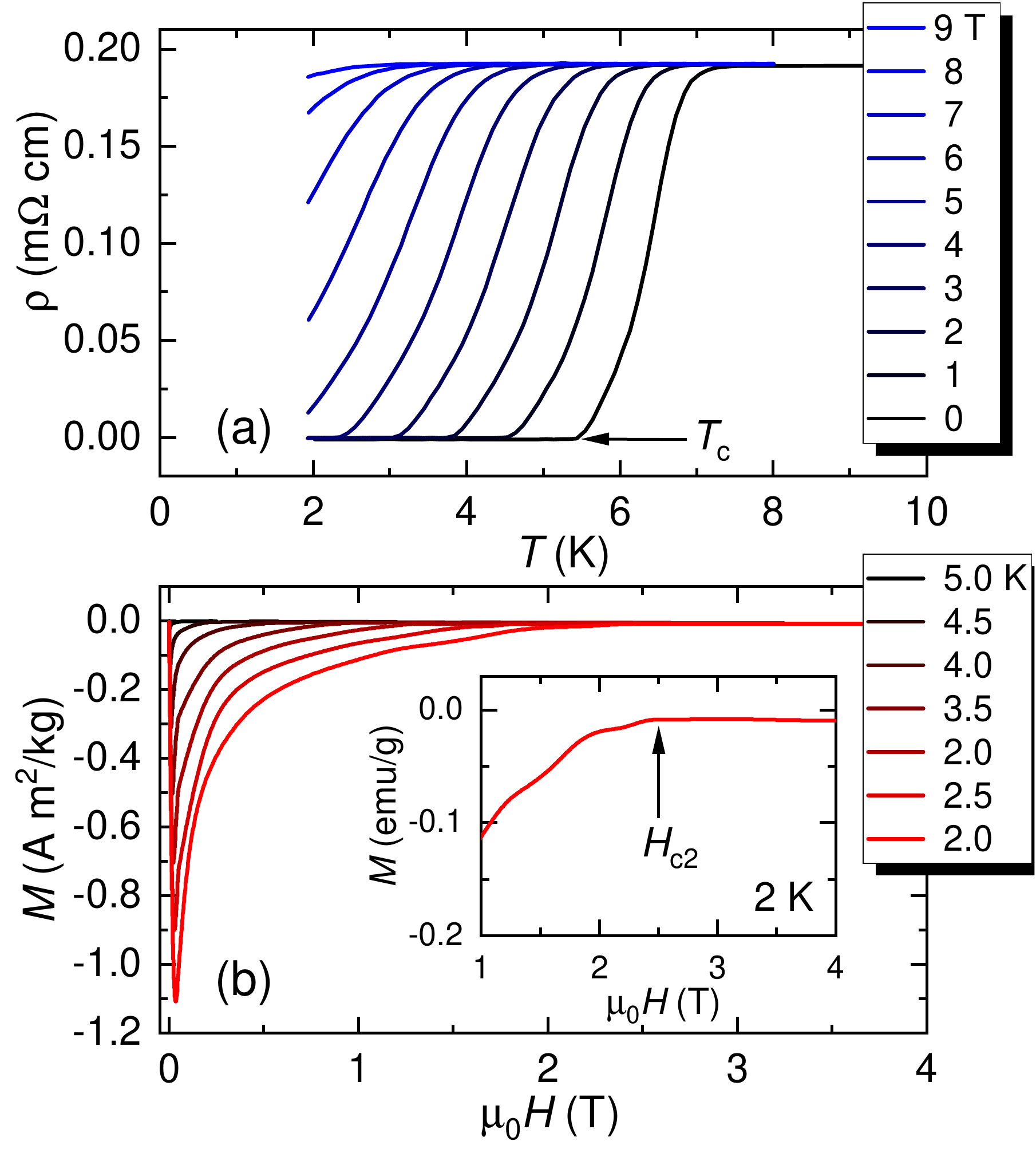}
	\caption{\label{fig:Hc2_raw}(a) Temperature-dependent electrical 
	resistivity for various applied magnetic fields.  
	 $T_c$ was determined as the onset temperature where the resistivity drops to zero. 
	 (b) Field-dependent magnetization (up to 4\,T) collected at various temperatures
	 below $T_c$. The inset shows the high-field range of the $M(H)$ curve at 2\,K.
	 $H_\mathrm{c2}$ was chosen as the field where the diamagnetic 
	 response vanishes 
	 (indicated by an arrow). 
	The reported data refer to the sample S2 --- samples S1 and S3 
	show similar features.}
\end{figure}
a clear diamagnetic response appears below the superconducting transition
at $T_c$ = 5.5\,K for S2. The samples S1 and S3 show a slightly lower 
transition temperature, i.e., $T_c$ $\sim$ 5.0\,K.
After accounting for the demagnetization factor, the superconducting 
shielding fraction of TaReSi samples is close to 100\%, indicative of bulk SC.

The temperature-dependent electrical resistivity $\rho(T)$ of TaReSi 
samples was measured from 2\,K up to room temperature. It reveals 
a metallic behavior, without 
any anomalies associated with structural, magnetic, or charge-density-wave 
transitions at temperatures above $T_c$ [see lower inset in Fig.~\ref{fig:Tc}(b)].
The electrical resistivity in the low-$T$ region is plotted in 
Fig.~\ref{fig:Tc}(b), clearly showing the superconducting transition
of all the samples. 
A $T^\mathrm{onset}_c = 6.1$, 7.0, and 7.5\,K, and 
$T^\mathrm{zero}_c = 4.8$, 5.6, and 4.6\,K were 
identified for the S1, S2, and S3 samples, respectively.
The $T^\mathrm{zero}_c$ values are consistent with 
the transition temperatures determined from the magnetic susceptibility 
[see Fig.~\ref{fig:Tc}(a)]. In view of its
higher $T_c$ 
and narrower $\Delta T_c$ transition, 
most of the {\textmu}SR measurements were performed on the TaReSi sample S2.  

To determine the lower critical field $H_\mathrm{c1}$, 
to be exceeded (at least twice)
when performing {\textmu}SR measurements on type-II superconductors, 
the field-dependent magnetization $M(H)$ of TaReSi was measured at 
various temperatures. Here, the $M(H)$ data of the S2 sample
are shown 
in Fig.~\ref{fig:Hc1}(a), with the other samples showing a similar behavior.
The estimated $H_\mathrm{c1}$ values at different temperatures 
(accounting for a demagnetization factor), determined 
from the deviations of $M(H)$ from linearity, are summarized in Fig.~\ref{fig:Hc1}(b).
The solid lines are fits to $\mu_{0}H_\mathrm{c1}(T) =\mu_{0}H_\mathrm{c1}(0)[1-(T/T_{c})^2]$ 
and yield the lower critical fields $\mu_{0}H_\mathrm{c1}(0)$ = 6.6(1), 6.4(1), and 5.5(1)\,mT for 
S1, S2, and S3 samples, respectively.

\begin{figure}[!htp]
	\centering
	\includegraphics[width=0.48\textwidth,angle= 0]{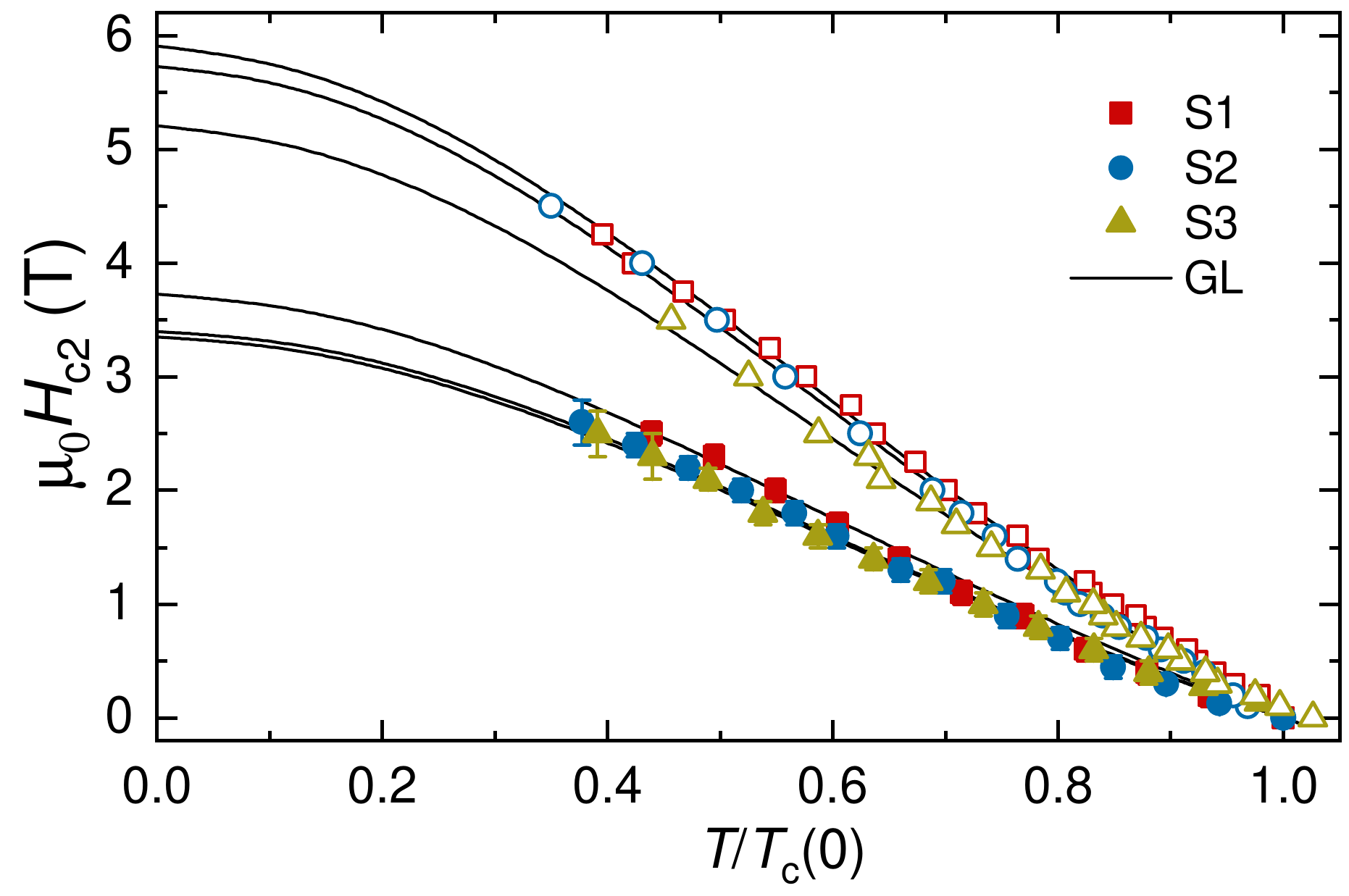}
	\caption{\label{fig:Hc2} Upper critical field $H_\mathrm{c2}$ vs.\ 
	the reduced 
    temperature $T/T_c(0)$ for 
	the different 
	TaReSi samples. The $T_c$ and $H_\mathrm{c2}$ values were determined from 
	the measurements shown in Fig.~\ref{fig:Hc2_raw}. 
    Full symbols refer to magnetization, while empty symbols to 
    resistivity measurements. 
    Note the systematically higher values in the latter case.
	Solid lines represent fits to the GL model.}
\end{figure}
To investigate the upper critical field $H_\mathrm{c2}$ of TaReSi, we 
measured the temperature-dependent electrical resistivity $\rho(T,H)$ 
at various applied magnetic fields, as well as the field-dependent 
magnetization $M(H,T)$ at various temperatures. 
As shown in Fig.~\ref{fig:Hc2_raw}(a), upon increasing the magnetic 
field, the superconducting transition in $\rho(T)$ 
shifts to lower temperatures. 
Similarly, in the $M(H)$ data, the diamagnetic signal vanishes 
once the applied magnetic field exceeds the upper critical field 
$H_\mathrm{c2}$ [see inset in Fig.~\ref{fig:Hc2_raw}(b)].
Figure~\ref{fig:Hc2} summarizes the upper critical fields $H_\mathrm{c2}$ 
vs.\ the reduced superconducting transition temperatures $T_c/T_c(0)$ for 
all the TaReSi samples, as identified
from the $\rho(T,H)$ and $M(H,T)$ data. 
To determine the upper critical field at 0\,K, the $H_\mathrm{c2}(T)$ 
data were analyzed by means of a semiempirical Ginzburg-Landau 
(GL) model,
$H_{c2} = H_{c2}(0)(1-t^2)/(1+t^2)$, where $t = T/T_c(0)$. 
As shown by the solid lines, the GL model gives 
$\mu_0$$H_\mathrm{c2}(0) = 3.7(1)$, 3.4(1), and 3.3(1)\,T for the  
TaReSi samples S1, S2, and S3, respectively. 
As for the electrical-resistivity data, the derived $H_\mathrm{c2}(0)$ 
values are much larger than the bulk values determined from magnetization data.  
The different $T_c$ or $H_\mathrm{c2}$ values might be related to 
a strongly anisotropic
upper critical field, or to the appearance of surface/filamentary 
superconductivity above bulk $T_c$.
Moreover, although the magnetization- and electrical-resistivity
measurements reveal different sample qualities, the superconducting
properties of TaReSi seem to be robust. Such an insensitivity of SC to
nonmagnetic impurities or disorder implies an $s$-wave pairing in
TaReSi, as  further evidenced by the {\textmu}SR measurements (see below).

In the GL theory of superconductivity, the coherence length 
$\xi$ can be calculated from $\xi$ =  $\sqrt{\Phi_0/2\pi\,H_{c2}}$, where 
$\Phi_0 = 2.07 \times 10^{3}$\,T~nm$^{2}$ is the quantum of
magnetic flux. 
With a bulk $\mu_{0}H_{c2}(0) = 3.4(1)$\,T (for S2 sample), the calculated $\xi(0)$ 
is 9.8(1)\,nm. The magnetic penetration depth $\lambda$ is related to 
the coherence length $\xi$ and the lower critical field $\mu_{0}H_{c1}$ 
via $\mu_{0}H_{c1} = (\Phi_0 /4 \pi \lambda^2)[$ln$(\kappa)+ 0.5]$, where 
$\kappa$ = $\lambda$/$\xi$ is the GL parameter~\cite{Brandt2003}.
By using $\lambda_0$ = 562(3)\,nm, we find $\mu_{0}H_{c1} = 2.4(1)$\,mT, 
which is smaller than the value determined from the magnetization data (see Fig.~\ref{fig:Hc1}).
Such difference in $H_\mathrm{c1}$, as well as the unusual behavior in $H_\mathrm{c2}$  
might be attributed to the anisotropic TaReSi superconductivity.
To clarify this, studies on the single crystals are required.

\begin{figure}[t]
	\centering
	\includegraphics[width=0.48\textwidth,angle= 0]{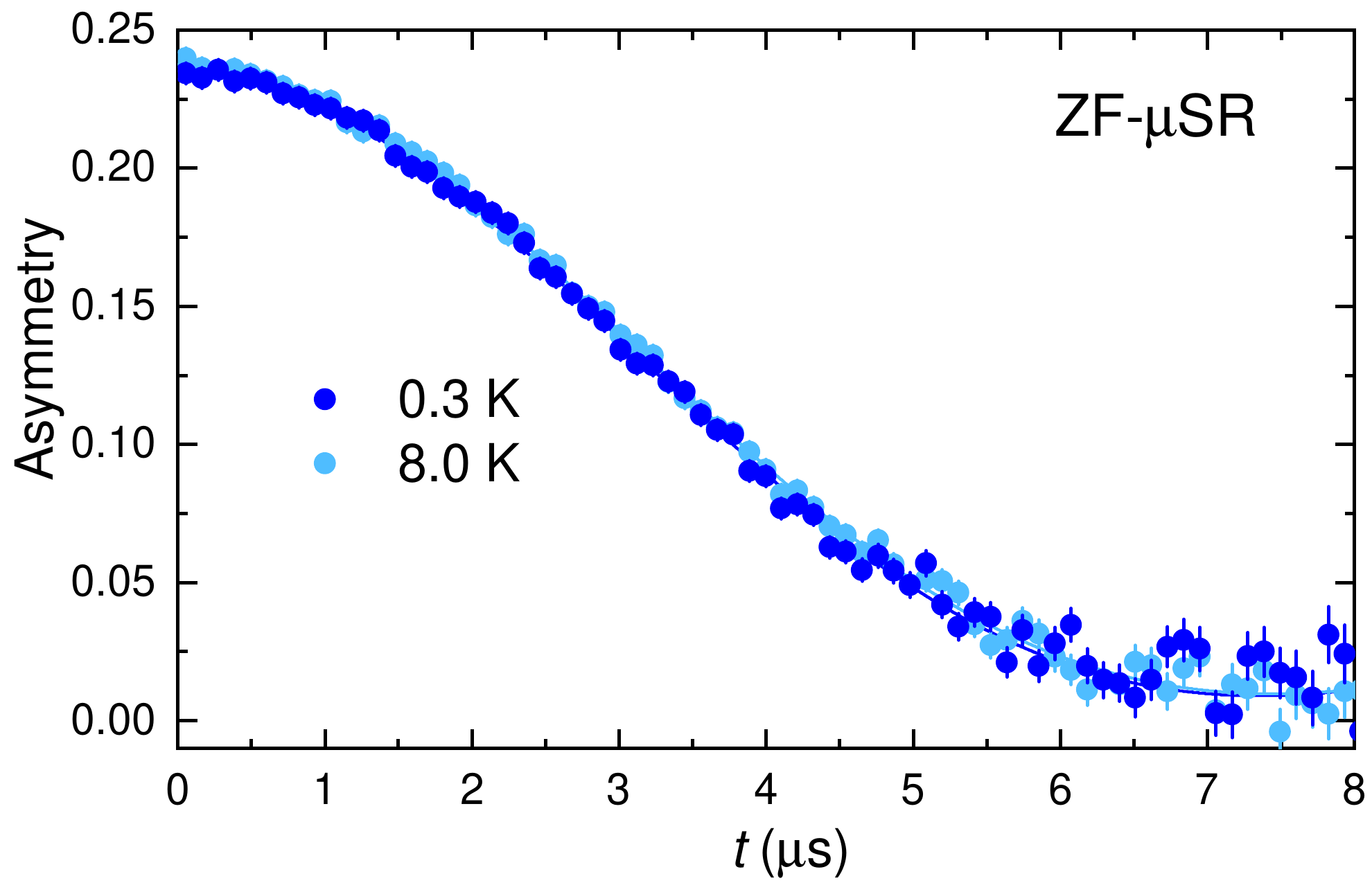}
	\caption{\label{fig:ZF-muSR}ZF-{\textmu}SR spectra collected in the 
	superconducting- (0.3\,K) and in the normal state (8\,K) 
	of TaReSi. 
	The practically overlapping datasets indicate the absence of 
	TRS breaking, whose occurrence would have resulted in a stronger 
	decay in the 0.3-K case.}
\end{figure}

Since certain
rhenium-based superconductors are known to break time-reversal 
symmetry in their superconducting state~\cite{Singh2014,Shang2018a,Shang2018b,Shang2020ReMo,Shang2021a}, 
to verify the possible breaking of TRS in TaReSi, we performed 
zero-field (ZF-) {\textmu}SR measurements in its normal- and superconducting states. 
This technique is very sensitive to the weak spontaneous fields 
expected to arise in these cases~\cite{Shang2021a}.
As shown in Fig.~\ref{fig:ZF-muSR}, the ZF-{\textmu}SR spectra 
of TaReSi lack any of the features associated with
magnetic order or magnetic fluctuations. Indeed, in the datasets
collected above- (8\,K) and below $T_c$ (0.3\,K), neither coherent 
oscillations nor fast decays could be identified. 
In the absence of an external magnetic field, the muon-spin relaxation
is mainly determined by the randomly oriented nuclear moments.
As a consequence, the ZF-{\textmu}SR spectra can be modeled by means of a 
phenomenological relaxation function, consisting of a combination of 
a Gaussian- and a Lorentzian Kubo-Toyabe relaxation~\cite{Kubo1967,Yaouanc2011}, i.e., 
$A_\mathrm{ZF} = A_\mathrm{s}[\frac{1}{3} + \frac{2}{3}(1 - 
\sigma_\mathrm{ZF}^{2}t^{2} - \Lambda_\mathrm{ZF} t)\,
\mathrm{e}^{(-\frac{\sigma_\mathrm{ZF}^{2}t^{2}}{2} - \Lambda_\mathrm{ZF} t)}] + A_\mathrm{bg}$.
Here,  $A_\mathrm{s}$ and $A_\mathrm{bg}$ are the initial asymmetries 
for the sample and sample holder, and $\sigma_\mathrm{ZF}$ and 
$\Lambda_\mathrm{ZF}$ represent the zero-field Gaussian and Lorentzian 
relaxation rates, respectively. The solid lines in Fig.~\ref{fig:ZF-muSR} 
are fits to the above equation, yielding $\sigma_\mathrm{ZF} = 0.231(1)$\,{\textmu}s$^{-1}$ 
and $\Lambda_\mathrm{ZF} = 0.005(2)$\,{\textmu}s$^{-1}$ at 8\,K and 
$\sigma_\mathrm{ZF} = 0.234(1)$\,{\textmu}s$^{-1}$ and $\Lambda_\mathrm{ZF} = 0.003(2)$\,{\textmu}s$^{-1}$ 
at 0.3\,K, respectively. The relaxation rates in the normal- and the 
superconducting states of TaReSi are almost identical, visually
confirmed by the
overlapping ZF-{\textmu}SR spectra in Fig.~\ref{fig:ZF-muSR}.
The absence of an additional {\textmu}SR relaxation below $T_c$ excludes 
the breaking of TRS in the superconducting state of TaReSi. 
\tcr{On the contrary, the enhanced $\sigma_\mathrm{ZF}$ below $T_c$ in
TaRuSi and NbRuSi provides clear evidence of the occurrence of spontaneous
magnetic fields, which break the TRS at the superconducting transition~\cite{Shang2022b}.
Such a selective occurrence of TRS breaking, 
observed also in other superconducting families~\cite{Shang2021a},
independent of ASOC, is puzzling and not yet fully understood,
clearly demanding further investigations. 
Future ZF-{\textmu}SR measurements on the TaRu$_{1-x}$Re$_{x}$Si series
could potentially clarify this issue. 
}

%
%
\begin{figure}[!thp]
	\centering
	\includegraphics[width=0.47\textwidth,angle= 0]{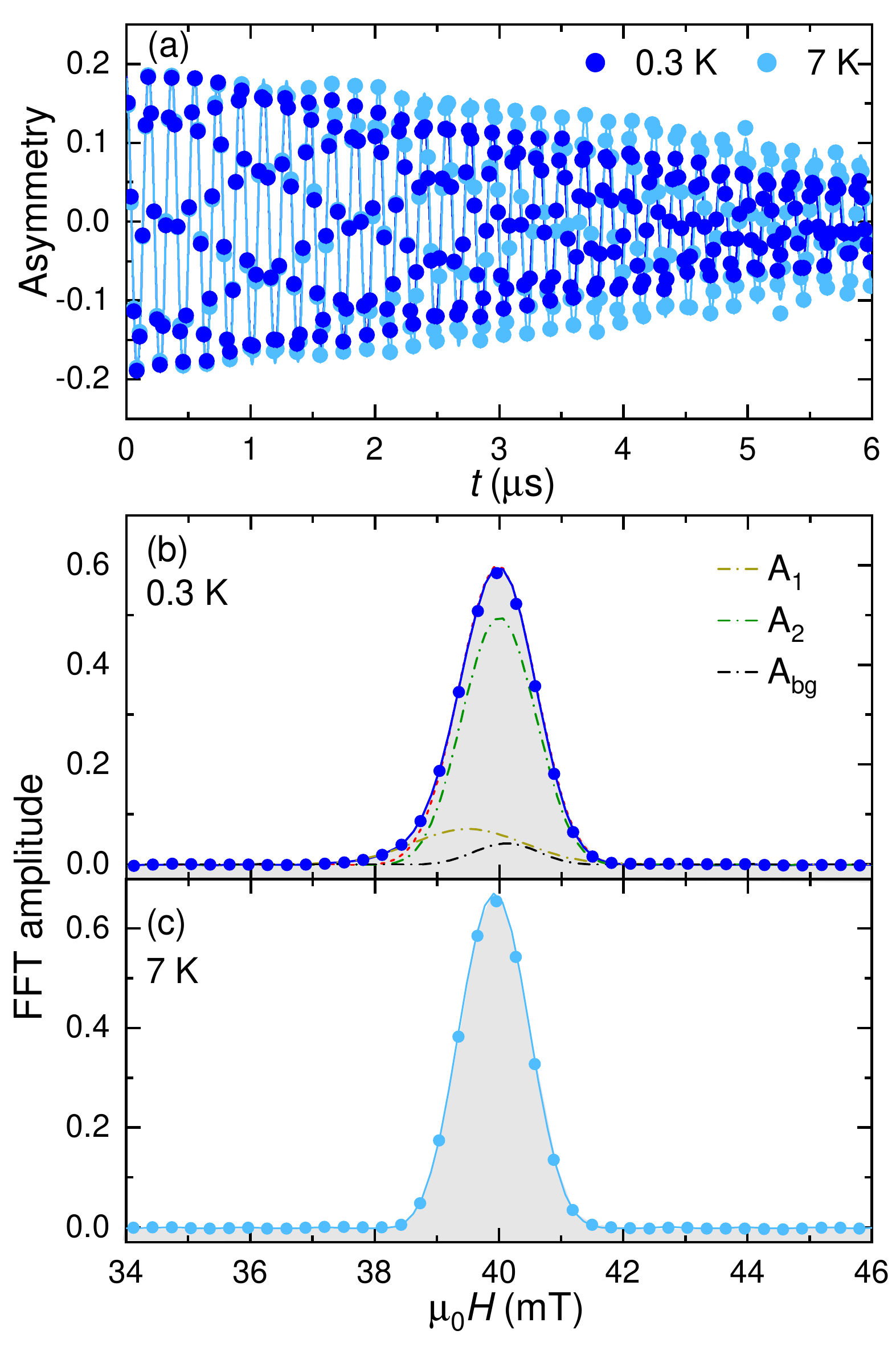}
	\caption{\label{fig:lambda}
	 (a) TF-{\textmu}SR spectra of TaReSi collected in the superconducting
	 (0.3\,K)- and normal (7\,K) states in an applied magnetic field
	 of 40\,mT. Dashed- and solid-lines are fits to Eq.~\eqref{eq:TF_muSR} using one- and two oscillations. 
   	 In the latter case, each contribution is shown separately as dash-dotted lines, together with a background contribution.
     Fits with two oscillations show a goodness-of-fit value $\chi_\mathrm{r}^2 \sim 1.0$, 
     smaller than the one-oscillation fits ($\chi_\mathrm{r}^2$ $\sim$ 1.6).}
\end{figure}
%
%
To investigate the superconducting pairing in TaReSi, we carried 
out systematic temperature-dependent transverse-field (TF-) {\textmu}SR 
measurements in an applied field of 40\,mT. 
Representative TF-{\textmu}SR spectra collected in the superconducting- and 
normal states of TaReSi are shown in Fig.~\ref{fig:lambda}(a). In the 
superconducting state (e.g., at 0.3\,K), the development of a 
flux-line lattice (FLL) causes an inhomogeneous field distribution 
and, thus, it gives rise to an additional damping in the 
TF-{\textmu}SR spectra~\cite{Yaouanc2011}. 
In such case, 
the TF-{\textmu}SR spectra are generally 
modeled using~\cite{Maisuradze2009}:
\begin{equation}
	\label{eq:TF_muSR}
	A_\mathrm{TF}(t) = \sum\limits_{i=1}^n A_i \cos(\gamma_{\mu} B_i t + \phi) e^{- \sigma_i^2 t^2/2} +
	A_\mathrm{bg} \cos(\gamma_{\mu} B_\mathrm{bg} t + \phi).
\end{equation}
%
%
%
%
%
\begin{figure}[!thp]
	\centering
	\includegraphics[width=0.48\textwidth,angle= 0]{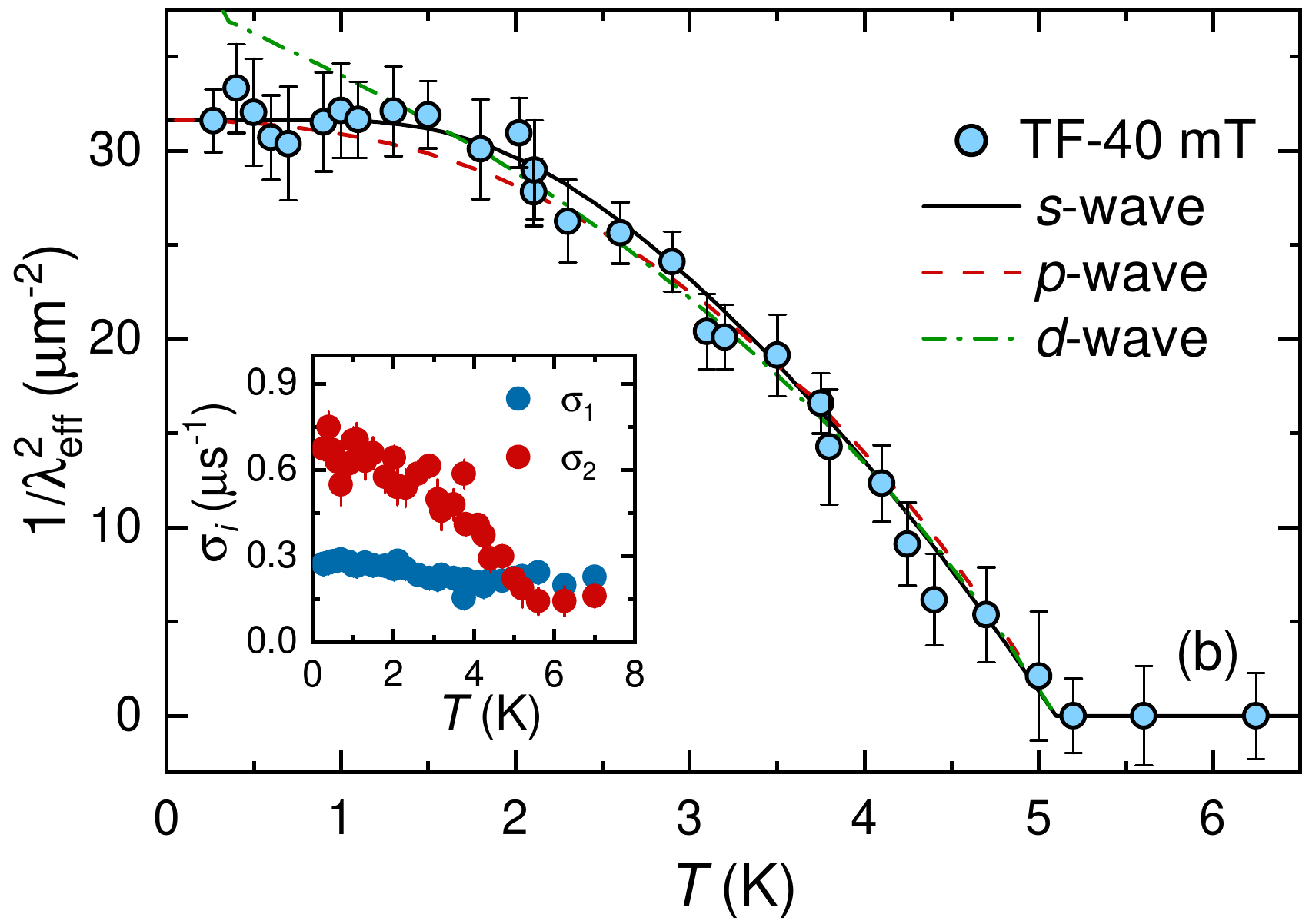}
	\caption{\label{fig:lambda2}%
	Temperature dependence of the su\-per\-fluid density 
	of TaReSi. The inset shows the muon-spin relaxation rates 
	$\sigma_i(T)$ vs.\ temperature. 
	The solid, dashed, and dash-dotted lines represent fits to the 
	$s$-, $p$-, and $d$-wave model, with $\chi_\mathrm{r}^2 \sim 1.1$,
	1.8, and 5.2, respectively.}
\end{figure}
%
%
Here $A_{i}$, $A_\mathrm{bg}$ and $B_{i}$, $B_\mathrm{bg}$ 
are the initial asymmetries and local fields sensed by implanted muons in the 
sample and sample holder,
$\gamma_{\mu}$/2$\pi$ = 135.53\,MHz/T 
is the muon gyromagnetic ratio, $\phi$ is a shared initial phase, and $\sigma_{i}$ 
is the Gaussian relaxation rate of the $i$th component. In general, the field distribution $p(B)$ in the superconducting state is material dependent.
In case of a symmetric $p(B)$, one oscillation (i.e., $n = 1$) 
is sufficient to describe the TF-{\textmu}SR spectra, while for an 
asymmetric $p(B)$, two or more oscillations (i.e., $n \ge 2$) are required.
Here, we find that Eq.~\eqref{eq:TF_muSR} with $n = 2$ can describe the experimental data quite well [see solid lines in Fig.~\ref{fig:lambda}(a)]. 
The derived muon-spin relaxation rates
$\sigma_i$ are small and 
temperature-independent in the normal state, but below $T_c$ they 
start to increase due to the onset of FLL and the increased superfluid 
density [see inset in Fig.~\ref{fig:lambda}(b)]. Then, the effective 
Gaussian relaxation rate $\sigma_\mathrm{eff}$ can be calculated
from $\sigma_\mathrm{eff}^2/\gamma_\mu^2 = \sum_{i=1}^2 A_i [\sigma_i^2/\gamma_{\mu}^2 - \left(B_i - \langle B \rangle\right)^2]/A_\mathrm{tot}$~\cite{Maisuradze2009}, where  $\langle B \rangle = (A_1\,B_1 + A_2\,B_2)/A_\mathrm{tot}$ and $A_\mathrm{tot} = A_1 + A_2$.
Considering the constant nuclear relaxation rate $\sigma_\mathrm{n}$ 
in the narrow temperature range investigated here, confirmed also 
by ZF-{\textmu}SR measurements (see Fig.~\ref{fig:ZF-muSR}), 
the superconducting Gaussian relaxation rate can be extracted using 
$\sigma_\mathrm{sc} = \sqrt{\sigma_\mathrm{eff}^{2} - \sigma^{2}_\mathrm{n}}$.

In TaReSi, the upper critical field $\mu_0H_\mathrm{c2}(0)$ $\sim$ 3.4\,T is significantly larger than the applied TF field (40\,mT). Hence, we can ignore
the effects of the overlapping vortex cores 
when extracting the magnetic penetration depth from the measured 
$\sigma_\mathrm{sc}$. The effective magnetic penetration depth 
$\lambda_\mathrm{eff}$ can then be calculated by using 
$\sigma_\mathrm{sc}^2(T)/\gamma^2_{\mu} = 0.00371\Phi_0^2/\lambda_\mathrm{eff}^4(T)$~\cite{Barford1988,Brandt2003}.
Figure~\ref{fig:lambda}(b) summarizes the temperature-dependent inverse 
square of magnetic penetration depth, which is proportional to the 
superfluid density, i.e., $\lambda_\mathrm{eff}^{-2}(T) \propto \rho_\mathrm{sc}(T)$.
The $\rho_\mathrm{sc}(T)$ was analyzed by applying different models, generally described by:

\begin{equation}
	\label{eq:rhos}
	\rho_\mathrm{sc}(T) = 1 + 2\, \Bigg{\langle} \int^{\infty}_{\Delta_\mathrm{k}} \frac{E}{\sqrt{E^2-\Delta_\mathrm{k}^2}} \frac{\partial f}{\partial E} \mathrm{d}E \Bigg{\rangle}_\mathrm{FS}. 
\end{equation}
Here, $f = (1+e^{E/k_\mathrm{B}T})^{-1}$ is the Fermi function and $\langle \rangle_\mathrm{FS}$ represents an average over the Fermi surface
(assumed to be an isotropic sphere, for $s$-wave superconductors)~\cite{Tinkham1996};
$\Delta_\mathrm{k}(T) = \Delta(T) \delta_\mathrm{k}$ is an angle-dependent
gap function, where $\Delta$ is the maximum gap value and $\delta_\mathrm{k}$ is the 
angular dependence of the gap, equal to 1, $\cos2\phi$, and $\sin\theta$ 
for an $s$-, $d$-, and $p$-wave model, respectively, with $\phi$ 
and $\theta$ being the azimuthal angles.
The temperature dependence of the gap is assumed to follow $\Delta(T) = \Delta_0 \mathrm{tanh} \{1.82[1.018(T_\mathrm{c}/T-1)]^{0.51} \}$~\cite{Tinkham1996,Carrington2003}, where $\Delta_0$ is the gap value at 0\,K.
Three different models, including $s$-, $p$-, and $d$-wave, were used to describe the $\lambda_\mathrm{eff}^{-2}$$(T)$ 
data. For an $s$- or $p$-wave model [see solid and dashed lines in Fig.~\ref{fig:lambda}(b)], the best fits yield the same 
zero-temperature magnetic penetration depth $\lambda_\mathrm{0} =562(3)$\,nm, but different superconducting gaps $\Delta_0$ =  0.79(2) and 1.05(2)\,meV, respectively. 
While for the $d$-wave model, the gap size is the same as $p$-wave model, but the $\lambda_\mathrm{0}$ = 510(3)\,nm is much shorter. 
As can be seen
in Fig.~\ref{fig:lambda}(b), the temperature-independent 
$\lambda_\mathrm{eff}^{-2}(T)$ for $T < 2$\,K strongly suggests a 
fully-gapped superconducting state 
in TaReSi. As a consequence, $\lambda_\mathrm{eff}^{-2}(T)$ is 
more consistent with the $s$-wave model, here reflected in the
smallest $\chi_\mathrm{r}^2$. In the case of a 
$p$- or $d$-wave model, a less-good agreement with the measured 
$\lambda_\mathrm{eff}^{-2}(T)$ is found, especially at low temperatures. 
\tcr{Although the unitary ($s + ip$) pairing~\cite{Shang2022b} can also 
describe the nodeless SC in TaReSi, its preserved TRS excludes such a possibility.
In summary,  TF-{\textmu}SR combined with ZF-{\textmu}SR data, 
indicate that TaReSi behaves as a conventional fully-gapped superconductor with preserved TRS.}

\begin{figure}[t]
	\centering{}
	\vspace{-1ex}
	\includegraphics[width=0.45\textwidth,angle=0]{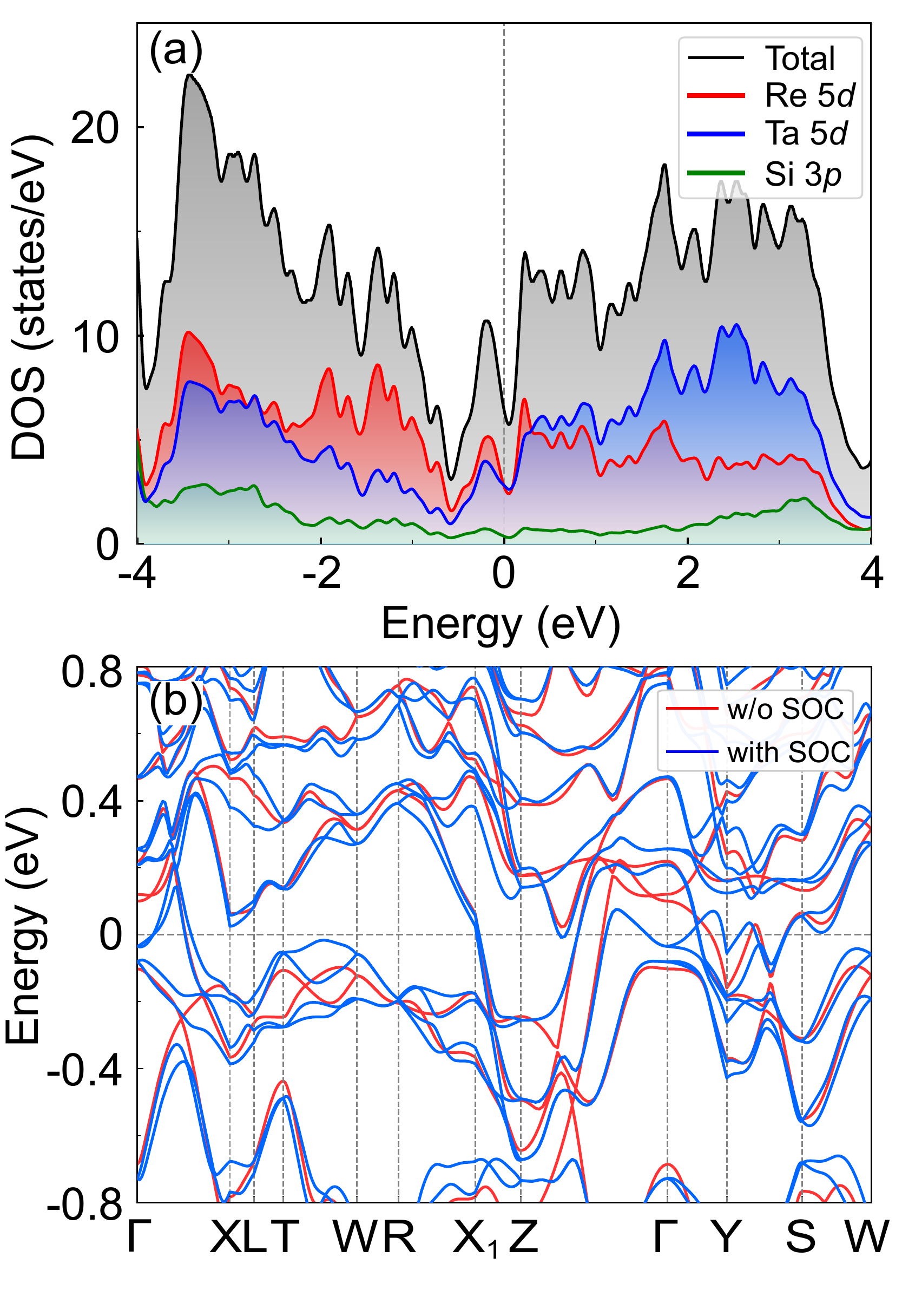}
	\caption{\label{fig:DOS}%
		(a) Calculated total- and partial (Ta-5$d$, Re-5$d$, and Si-3$p$ orbitals) 
		density of states for TaReSi. 
		(b) Electronic band structure of TaReSi, calculated by ignoring 
		(red) and by considering (blue) the spin-orbit coupling. 
		Several bands cross the Fermi level.}
\end{figure}
%

We also note that due to the lack of inversion symmetry in
TaReSi, a mixing of spin-singlet and spin-triplet pairing is \tcr{allowed}. 
Such mixing not only can be consistent with a fully-gapped
superconducting state but, more importanly, it can lead to
unconventional or even topological SC.  
Indeed, our TF-{\textmu}SR results clearly suggests a fully-gapped superconducting
state, here fitted by using an $s$-wave model (see Fig.~\ref{fig:lambda}). However, this does not imply that $s$-wave pairing is the only possibility. A mixed singlet-triplet pairings also allow a fully-gapped superconducting state, which in principle is allowed by the presence of ASOC~\cite{Hu2021}. 
Furthermore, topological \tcr{SC can occur}
when the pairing gap changes sign on different Fermi surfaces according to the topological criterion~\cite{Qi2010}. For a minimal single-band model, there are two spin-split Fermi surfaces, whose gaps are given by $\Delta_\mathrm{s}$ $\pm$  $\Delta_\mathrm{t}$($k_\mathrm{F}$), which implies that a sign change occurs when $\Delta_\mathrm{s}$ < $\Delta_\mathrm{t}$($k_\mathrm{F}$).

\begin{figure*}[t]
	\centering
	\includegraphics[width=0.9\textwidth,angle=0]{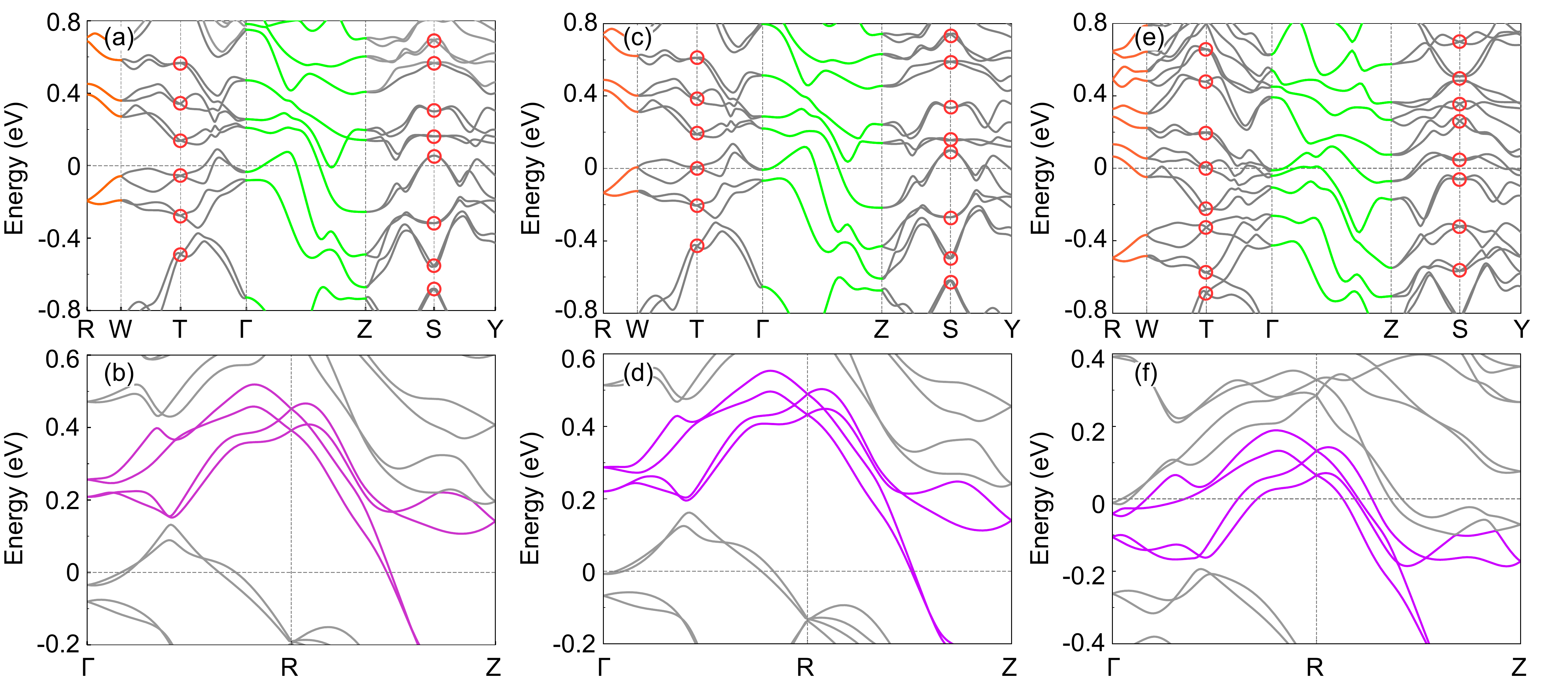}
	\vspace{-2ex}%
	\caption{\label{fig:kramers}(a) Illustration
			of Kramers Weyl points and Kramers nodal lines in TaReSi. 
			The KWP are marked by red circles, while the KNL are depicted
			by green- (along $\Gamma$--$Z$) or orange lines (along $R$--$W$), respectively. 
			(b) Illustration 
			of hourglass-shaped dispersion (purple lines) for TaReSi along 
			the $\Gamma$--$R$--$Z$ direction. The analogous results for TaReSi doped with 5\% Hf and 50\% W are shown in panels (c)-(d) and (e)-(f), respectively. 
			}
\end{figure*}
%

To gain further insight into the electronic properties of the
TaReSi superconductor, we also performed band-structure calculations 
using the density-functional theory.  
The electronic band structure of TaReSi, as well as its density 
of states (DOS) are summarized in Fig.~\ref{fig:DOS}. 
Close to the Fermi level $E_\mathrm{F}$, the DOS is 
dominated by the Ta- and Re-5$d$ orbitals, while the contribution 
from Si-3$p$ orbitals is negligible. 
The dominance of high-$Z$ orbitals might lead to a 
relatively large band splitting.
In TaReSi, the estimated DOS at $E_\mathrm{F}$ is about 
1.1 states/(eV f.u.) [= 6.5 states/(eV cell)/$Z$, 
with $Z = 6$ the number of atoms per primitive cell]. This is
comparable to the experimental value of 2.3 states/(eV f.u.), 
determined from the electronic specific-heat coefficient~\cite{Sajilesh2021}. 
The electronic band structure  
of TaReSi, calculated by ignoring and by considering the spin-orbit coupling, 
is shown in Fig.~\ref{fig:DOS}(b). 
When taking SOC into account, the electronic bands split 
due to the lifting of degeneracy, with one of them ending up closer to 
the Fermi level. The band splitting $E_\mathrm{ASOC}$ caused by the 
antisymmetric spin-orbit coupling is clearly visible in TaReSi, e.g., 
near the $X$ ($X_1$), $Y$, and $W$ points.  
The estimated band splitting in TaReSi is 
$E_\mathrm{ASOC} \sim 300$\,meV, 
\tcr{which is much larger than that of NbRuSi ($\sim$ 100\,meV),
but comparable to TaRuSi ($\sim$300\,meV)~\cite{Shang2022b}.}
Though smaller than the band splitting in CePt$_3$Si~\cite{Samokhin2004}, 
it is comparable to that of CaPtAs and Li$_2$Pt$_3$B~\cite{Shang2020,Lee2005}, 
and much larger than that of most other weakly-correlated NCSCs~\cite{Smidman2017}.
The $E_\mathrm{ASOC}$ of TaReSi is 
almost twice larger than that of the
analog NbReSi compound ($E_\mathrm{ASOC}$ $\sim$ 150\,meV)~\cite{Su2021,Shang2022}. 
The latter crystallizes in a ZrNiAl-type noncentrosymmetric structure 
($P\overline{6}2m$, No.~189) and exhibits features of 
unconventional superconductivity, 
e.g., its $H_\mathrm{c2}$ exceeds the Pauli limit. 
However, the $H_\mathrm{c2}$ of TaReSi is much smaller than that of 
NbReSi, the former being mostly determined by the orbital limit.  
Since Ta has a much larger atomic number than Nb (and, hence, a 
larger SOC), it is not surprising that TaReSi exhibits a larger 
$E_\mathrm{ASOC}$, in particular, considering that its Ta-$5d$
(instead of Nb-$4d$) orbitals contribute as much as Re-$5d$
orbitals to the DOS at the Fermi energy [see Fig.~\ref{fig:DOS}(a)].

According to the topological-materials database~\cite{Bradlyn2017,Vergniory2019,Zhang2019,Tang2019,TMD,BCS} and from 
our own band-structure calculations, 
TaReSi can be classified as a symmetry-enforced semimetal, \tcr{which
shares a similar band topology with NbRuSi and TaRuSi~\cite{Shang2022b}}. 
In the presence of spin-orbit coupling, owing to its nonsymmorphic 
space group ($Ima2$, No.~46), TaReSi 
hosts Kramers Weyl points 
(KWP) at the high-symmetry points and Kramers nodal lines along 
the high-symmetry lines of its Brillouin zone.
These features are marked by red circles (KWP) and green/orange 
lines (KNL) in Fig.~\ref{fig:kramers}(a).
The high-symmetry points at $S$ and $T$ are time-reversal symmetry invariant. 
As a consequence, the respective energies exhibit a twofold Kramers degeneracy 
protected by TRS. At the same time, due to the lack of inversion symmetry in TaReSi, these 
points cannot achieve the fourfold degeneracy of Dirac points and, hence, they are Weyl points.
As for the high-symmetry lines along the $\Gamma$-$Z$ and $R$-$W$ 
directions, the bands form a two-dimensional representation, 
i.e., twofold degenerate, indicating the occurrence of KNL 
in TaReSi.
Since most of KNLs occur near the $E_\mathrm{F}$, with a few of them even
crossing it, \tcr{similarly to NbRuSi and TaRuSi~\cite{Shang2022b},
TaReSi can be classified as a Kramers nodal-line semimetal (KNLS).
At the high-symmetry $S$ and $T$ points, the KWP in TaReSi are closer
to $E_\mathrm{F}$ than in NbRuSi and TaRuSi~\cite{Shang2022b}.
Since the Ru atoms have one more electron than Re, the KWP in NbRuSi
and TaRuSi is shifted further below $E_\mathrm{F}$.}

More interestingly, as shown by purple lines in Fig.~\ref{fig:kramers}(b), 
due to its nonsymmorphic space-group symmetry, TaReSi also exhibits 3D 
bulk hourglass-type fermions, characterized by an hourglass cone with five 
doubly degenerate points~\cite{Wang2016,Wu2020}. The $Ima2$ nonsymmorphic 
space group contains the generator of a glide mirror reflection  
$M_y=\{m_{010}\vert 1/2,0,0\}$: $(x,y,z)$$\to$$(x+1/2,-y,z)$ [see inset in Fig.~\ref{fig:Tc}(a)]. 
Here, the $k_y = 0$- and $\pi$-planes are $M_y$-invariant planes, 
where all the states along the $\Gamma$--$R$--$Z$ line carry the 
$M_y$-index $\pm i e^{-ik_x/2}$ and give rise to the 3D bulk 
hourglass fermions, protected by the $M_y$ operator~\cite{Motohiko2016,Wang2017}. 
In this case, at a high-symmetry point, the {\bf k}-vectors are ${\bf k}_\Gamma=(0,0,0)$, ${\bf k}_R=(0,1/2,0)$ and ${\bf k}_Z=(1/2,1/2,-1/2)$. Therefore, the $M_y$-index is $\pm 1$ 
at the $R$ point, and $\pm i$ at $\Gamma$ and $Z$ points. 
In agreement with the Kramers theorem, each state is twofold 
degenerate, i.e., pairs of doubly degenerate states exhibit identical
energies, but carry opposite $M_y$-indexes.
In the presence of a strong SOC, these doubly degenerate states 
split along the $R\to\Gamma$ or $R\to Z$ directions.
Despite this SOC-induced splitting of bands with different $M_y$-indexes, 
a residual degeneracy remains, which could give rise to 
the non-interacting hourglass fermions in TaReSi.
To date, hourglass fermions were experimentally observed only in very 
few materials, as e.g., 
the KHgSb and Nb$_3$(Si,Ge)Te$_6$ 
topological insulators~\cite{Ma2017,Wan2021}. Here, we establish
that \tcr{similar to the NbRuSi and TaRuSi compounds~\cite{Shang2022b},}
 also TaReSi belongs to this restricted class of materials, where Kramers Weyl points and hourglass
	fermions exist and can be tuned toward $E_\mathrm{F}$ by Hf- or W- chemical substitutions on the Ta site  
[see Fig.~\ref{fig:kramers}(c)-(d) for 5\%-Hf substitution and Fig.~\ref{fig:kramers}(e)-(f) for 50\%-W substitution].
At the same time, we could show that neither chemical substitution
on the Si site (here introduced via Si-to-Ge substitution), nor physical
pressure have appreciable effects on the band structure of TaReSi.
Besides exhibiting nontrivial electronic bands, TaReSi shows 
also intrinsic SC at low temperatures. 
This remarkable combination makes it a promising candidate material
for investigating topological properties.

\section{\label{ssec:Sum}Conclusion}\enlargethispage{8pt}
To summarize, we studied the noncentrosymmetric TaReSi superconductor 
by means of {\textmu}SR measurements and band-structure calculations.  
The superconducting state of TaReSi is characterized by a $T_c$ of $\sim$5.5\,K and an upper critical field $\mu_0H_{c2}(0)$ of $\sim$3.4\,T.  
The temperature-dependent superfluid density reveals a 
\emph{fully-gapped} superconducting state in TaReSi.
The lack of spontaneous magnetic fields below $T_c$ indicates 
a \emph{preserved} time-reversal symmetry in the superconducting state of TaReSi. 
Electronic band-structure calculations reveal that TaReSi \tcr{shares
a similar band topology to NbRuSi and TaRuSi, which also} belong to the 
three-dimensional Kramers nodal-line semimetals. 
It, too, features hourglass fermions, protected by the nonsymmorphic 
space-group symmetry.
Our results demonstrate that TaReSi represents 
a potentially interesting 
system for investigating the rich interplay between the exotic electronic states of 
Kramers nodal-line fermions, hourglass fermions, and superconductivity.
It will be also interesting to explore the Zeeman-field-induced 
Weyl superconductor in this material.
Considering the nontrivial band structure near the Fermi level and its 
intrinsic superconductivity, TaReSi \tcr{represents one of the}
promising platforms for investigating the topological
aspects of noncentrosymmetric superconductors.

\vspace{1pt}
\begin{acknowledgments}
This work was supported by the Natural Science Foundation of Shanghai 
(Grants No.\ 21ZR1420500 and 21JC\-140\-2300), Natural Science Foundation of Chongqing (Grant No.\ CSTB2022NSCQ-MSX1678),  and the Schweizerische 
Nationalfonds zur F\"{o}r\-der\-ung der Wis\-sen\-schaft\-lichen For\-schung 
(SNF) (Grants No.\ 200021\_188706 and 206021\_139082). Y.X. acknowledges
support from the Shanghai Pujiang Program (Grant No. 21PJ1403100).
We acknowledge the allocation of beam time at the Swiss muon source 
(Dolly {\textmu}SR spectrometer).
\end{acknowledgments}


\bibliography{TaReSi.bib}

\end{document}